%% file: nmpf.tex
\documentclass[pre,twocolumn,showpacs,showkeys,superscriptaddress,preprintnumbers,floatfix]{revtex4-1}

\usepackage{etex}
\usepackage{ifpdf}
\usepackage{hyperref}
\usepackage{dcolumn}
\usepackage{url}
\usepackage{amsmath}
\usepackage{amscd}
\usepackage{amsfonts}
\usepackage{amssymb}
\usepackage{amsthm}
\usepackage{xfrac}
\usepackage{braket}
\usepackage{bm}   % bold math
\usepackage{bbm}
\usepackage{verbatim}
\usepackage{mathtools}
\usepackage{stmaryrd}
\usepackage{xcolor}
\usepackage{setspace}
\usepackage{tikz}
\usetikzlibrary{arrows}
\usetikzlibrary{automata}
\usetikzlibrary{calc}
\usetikzlibrary{decorations.pathreplacing}
\usetikzlibrary{patterns}
\usetikzlibrary{positioning}
\usetikzlibrary{plotmarks}
\usetikzlibrary{shapes}

\usepackage{pgfplots}
\pgfplotsset{compat=newest}

\tikzstyle{vaucanson}=[
  node distance=2.5cm,
  bend angle=15,
  auto,
  every loop/.style={},
  every edge/.style={->,draw=black,line width=1.2,>=latex,shorten <=1pt,shorten >=1pt},
  every state/.style={draw=black,line width=2,font=\large},
  loop right/.style={right,out=22,in=-22,loop},
  loop above/.style={above,out=112,in=68,loop},
  loop left/.style={left,out=202,in=158,loop},
  loop below/.style={below,out=292,in=248,loop},
  loop above right/.style={right,out=67,in=23,loop},
  loop above left/.style={left,out=157,in=113,loop},
  loop below left/.style={left,out=247,in=203,loop},
  loop below right/.style={right,out=337,in=293,loop},
]

\theoremstyle{plain}    
\theoremstyle{plain}    
\theoremstyle{plain}    
\theoremstyle{plain}    
\theoremstyle{plain}    
\theoremstyle{plain}    
\theoremstyle{plain}    
\theoremstyle{plain}    
\theoremstyle{plain}    
\theoremstyle{plain}    
\theoremstyle{plain}    
\theoremstyle{plain}    
\theoremstyle{plain}    
\theoremstyle{plain}    
\theoremstyle{plain}    
\theoremstyle{plain}    
\theoremstyle{plain}

\input{cmechabbrev}

\renewcommand{\H}{\operatorname{H}}
\renewcommand{\I}{\operatorname{I}}

\colorlet {R_color}    {blue}
\colorlet {k_color}    {black!30!green}

\def\clap#1{\hbox to 0pt{\hss#1\hss}}

\newcommand{\approxprop}{ \begin{subarray}{l} \propto \\[-2.0pt] \sim \end{subarray} }

\begin{document}

\title{Nearly Maximally Predictive Features and Their Dimensions}

\author{Sarah E. Marzen}
\email{semarzen@mit.edu}
\affiliation{Physics of Living Systems Group, Department of Physics, Massachusetts Institute of Technology, Cambridge, MA 02139}                     \affiliation{Department of Physics, University of California at Berkeley, Berkeley, CA 94720-5800}

\author{James P. Crutchfield}
\email{chaos@ucdavis.edu}
\affiliation{Complexity Sciences Center, Department of Physics\\
University of California at Davis, One Shields Avenue, Davis, CA 95616}

\date{\today}
\bibliographystyle{unsrt}

% ************************* ABSTRACT *************************
\begin{abstract}
Scientific explanation often requires inferring maximally predictive features
from a given data set. Unfortunately, the collection of minimal maximally
predictive features for most stochastic processes is uncountably infinite. In
such cases, one compromises and instead seeks nearly maximally predictive
features. Here, we derive upper-bounds on the rates at which the number and the
coding cost of nearly maximally predictive features scales with desired
predictive power. The rates are determined by the fractal dimensions of a
process' mixed-state distribution. These results, in turn, show how widely-used
finite-order Markov models can fail as predictors and that mixed-state
predictive features offer a substantial improvement.
\end{abstract}

\keywords{hidden Markov models, entropy rate, dimension, resource-prediction
trade-off, mixed-state simplex}

\pacs{
02.50.-r  %  Probability theory, stochastic processes, and statistics
89.70.+c  %  Information science
05.45.Tp  %  Time series analysis
% 02.50.Ey  %  Stochastic processes
02.50.Ga  %  Markov processes
% 05.20.-y  %  Classical statistical mechanics
% 05.45.-a  %  Nonlinear dynamics and nonlinear dynamical systems
% 89.75.Kd  %  Complex Systems: Patterns
}
\preprint{Santa Fe Institute Working Paper 17-02-XXX}
\preprint{arxiv.org:1702.XXXX [physics.gen-ph]}

\maketitle

% ****************************************************************

%\tableofcontents
\setstretch{1.1}

% Handy abbreviations in the following
\newcommand{\Abet}{\ProcessAlphabet}
\newcommand{\MS}{\MeasSymbol}
\newcommand{\ms}{\meassymbol}
\newcommand{\SSet}{\CausalStateSet}
\newcommand{\St}{\CausalState}
\newcommand{\st}{\causalstate}
\newcommand{\MxSt}{\AlternateState}
\newcommand{\MxSSet}{\AlternateStateSet}
\newcommand{\mxst}{\mu}
\newcommand{\mxstt}[1]{\mu_{#1}}
\newcommand{\StartMS}{\bra{\delta_\pi}}
\newcommand{\Ipred}{\EE}
\newcommand{\ISI} { \xi }

\newcommand{\ECT}{\widehat{\EE}}
\newcommand{\CCT}{\widehat{C}_\mu}

\newcommand{\gen}{g}
\newcommand{\FeatAlphabet}{\mathcal{F}}

%%%%%%%%%%%%%%%%%%%%%%%%%%%%%%%%%%%%%%%%%%

%\vspace{0.2in}
%\section{Introduction}

Often, we wish to find a minimal maximally predictive model consistent with
available data. Perhaps we are designing interactive agents that reap greater
rewards by developing a predictive model of their environment \cite{Stil07c,
littman2001predictive, tishby2011information,Litt13a,Still2012, Brodu11} or,
perhaps, we wish to build a predictive model of experimental data because we
believe that the resultant model gives insight into the underlying mechanisms
of the system \cite{Crut92c, Stil07a}. Either way, we are almost always faced
with constraints that force us to efficiently compress our data \cite{Berg71a}.

%(\alert{Add schmidhuber, singh predictive states, recent tishby/polani})

Ideally, we would compress information about the past without sacrificing any
predictive power. For stochastic processes generated by finite unifilar hidden
Markov models (HMMs), one need only store a finite number of predictive
features. The minimal such features are called \emph{causal states}, their
coding cost is the \emph{statistical complexity} $\Cmu$ \cite{Crut88a}, and the
implied unifilar HMM is the \emph{\eM} \cite{Crut88a,Shal98a}. However, most
processes require an infinite number of causal states \cite{Crut92c} and
so cannot be described by finite unifilar HMMs.

In these cases, we can only attain some maximal level of predictive power given constraints on the number of predictive features or their coding cost.  Equivalently, from finite data we can only infer a finite predictive model.
Thus, we need to know how our predictive power grows with available resources.

Recent work elucidated the tradeoffs between resource constraints and
predictive power for stochastic processes generated by countable unifilar
HMMs or, equivalently, described by a finite or countably infinite number of
causal states \cite{creutzig2009past,Stil07b,Marz14f}. Few, though, studied
this tradeoff or provided bounds thereof more generally.

Here, we place new bounds on resource-prediction tradeoffs in the limit of
nearly maximal predictive power for processes with either a countable or an
uncountable infinity of causal states by coarse-graining the \emph{mixed-state
simplex} \cite{Blac57b}. These bounds give a novel operational
interpretation to the fractal dimension of the mixed-state simplex and suggest
new routes towards quantifying the memory stored in a stochastic process when,
as is typical, statistical complexity diverges.

\paragraph*{Background}
We consider a discrete-time, discrete-state stochastic process $\Process$
generated by an HMM $\Gen$, which comes equipped with underlying states $\gen$
and labeled transition matrices $T^{\ms}_{\gen,\gen'} =
\Prob(\Gen_{t+1}=\gen',\MS_{t+1}=\ms|\Gen_t=\gen)$ \cite{Rabi86a}. There is an
infinite number of alternate HMMs that generate $\Process$
\cite{gilbert1959identifiability,Ito92a,Bala97a,Uppe97a}, so we specify here
that $\Gen$ is the \emph{minimal} generative model---that is, the generative
model with the minimal number of hidden states consistent with the observed
process \footnote{This is not necessarily the same as the generative model with
minimal generative complexity \cite{lohrthesis}, since a nonunifilar generator
with a large number of sparsely-used states might have smaller state entropy
than a nonunifilar generator with a small number of equally-used states. And,
typically, per our main result, the minimal generative model is usually not the
same as the minimal prescient (unifilar) model.}.

For reasons that become clear shortly, we are interested in the \emph{block
entropy} $\H(L) = \H[\MS_{0:L}]$, where $\MS_{a:b} = \MS_a, \MS_{a+1},
\ldots, \MS_{b-1}$ is a contiguous block of random variables generated by $\Gen$. In particular, its growth---the \emph{entropy
rate} $\hmu = \lim_{L\rightarrow\infty} \H(L) / L$---quantifies a process'
intrinsic ``randomness''. While the \emph{excess entropy} $\EE =
\lim_{L\rightarrow\infty} \left( \H(L)-\hmu L \right)$ quantifies how much is
predictable: how much future information can be predicted from the past
\footnote{Cf. Ref. \cite{Crut01a}'s discussion of terminology and Ref.
\cite{Bial00a}}. Finite-length entropy-rate estimates $\hmu(L) =
\H[\MS_0|\MS_{-L:0}]$ provide increasingly better approximations to the true
entropy rate $\hmu$ as $L$ grows large. While finite-length excess-entropy
estimates:
\begin{align}
\EE(L) & = \H(L)-\hmu L \\
       & = \sum_{\ell=0}^{L-1} \left( \hmu(\ell)-\hmu \right)
	   ~.
\label{eq:EEL}
\end{align}
tend to the true excess entropy $\EE$ as $L$ grows large \cite{Crut01a}. As
there, we consider only \emph{finitary} processes, those with finite $\EE$ \footnote{Finite HMMs generate finitary processes, as $\EE$ is bounded from above by the logarithm of the number of HMM states \cite{lohrthesis}}.

Predictive features $\Rep$ from some alphabet $\FeatAlphabet$ are formed by
compressing the process' past $\MS_{-\infty:0}$ in ways that implicitly retain
information about the future $\MS_{0:\infty}$. (From hereon, our block notation
suppresses infinite indices.) A \emph{predictive distortion} quantifies the
predictability lost after such a coarse-graining:
\begin{align*}
d(\Rep) & = \I[\Past;\Future|\Rep] \\
        & = \EE - \I[\Rep;\Future]
  ~.
\end{align*}
On the one hand, distortion achieves its maximum value $\EE$ when $\Rep$
captures no information from the past that could be used for prediction. On the
other, it can be made to vanish trivially by taking the predictive features
$\Rep$ to be all of the possible histories $\Past$. Here, we quantify the cost
of a larger feature space in two ways: by the number $|\FeatAlphabet|$ of
predictive features and by their coding cost $H[\Rep]$ \cite{Cove06a}.

%\alert{Things getting a little overloaded: predictive features,
%prescient states, causal states, feature alphabet, predictors v. generators.
%SM: not sure how to deal with this.}

\subparagraph*{Results}
Ideally, we would identify the minimal number $|\FeatAlphabet|$ of predictive
features or the coding cost $H[\Rep]$ required to achieve at least a given
level $d$ of predictive distortion. This is almost always a difficult
optimization problem. However, we can place upper bounds on $|\FeatAlphabet|$
and $H[\Rep]$ by constructing suboptimal predictive feature sets that achieve
predictive distortion $d$.

We start by reminding ourselves of the optimal solution in the limit that $d =
0$---the \emph{causal states} $\St$ \cite{Crut88a,Shal98a,Stil07b}. Causal
states can be defined by calling two pasts, $\past$ and $\past'$,
\emph{equivalent} when using them our predictions of the future are the same:
$\Prob(\Future|\Past=\past) = \Prob(\Future|\Past=\past')$. (These
conditional distributions are referred to as \emph{future morphs}.) One can then form a
model, the \emph{\eM}, from the set $\St$ of causal-state equivalence classes
and their transition operators. The Shannon entropy of the causal state
distribution is the \emph{statistical complexity}: $\Cmu = \H[\St]$. A process'
\eM\ can also be viewed as the minimal unifilar HMM capable of generating the
process \cite{Trav11a} and $\Cmu$ the amount of historical information the
process stores in its causal states. Though the mechanics of working with
causal states can become rather sophisticated, the essential idea is that
causal states are designed to capture everything about the past relevant to
predicting the future and \textit{only} that information.

In the lossless limit, when $d=0$, one cannot find predictive representations
that achieve $|\FeatAlphabet|$ smaller than $|\St|$ or that achieve $\H[\Rep]$
smaller than $\Cmu$. Similarly, no optimal lossy predictive representation will
ever find $|\FeatAlphabet|>|\St|$ or $\H[\Rep]\geq\Cmu$. When the number of
causal states is infinite or statistical complexity diverges, as is typical, as
we noted, these bounds are quite useless; but otherwise, they provide a useful
calibration for the feature sets proposed below.

\paragraph*{Markov Features}
Several familiar predictive models use pasts of length $L$ as predictive
features. This feature set can be thought of as constructing an order-$L$
Markov model of a process \footnote{One can construct a better feature set
simply by applying the causal-state equivalence relation to these length-$L$
pasts. We avoid this here since the size of this feature set is more difficult
to analyze generically and since the causal-state equivalence relation applied
to length-$L$ pasts often induces no coarse-graining.}. The implied predictive
distortion is $d(L) = \I[\Past;\Future|\MS_{0:L}] = \EE-\EE(L)$ \cite{Ara14a}, while
the number of features is the number of length-$L$ words with nonzero
probability and their entropy $\H[\Rep] = \H(L)$. Generally, $\hmu(L)$
converges exponentially quickly to the true entropy rate $\hmu$ for stochastic
processes generated by finite-state HMMs, unifilar or not; see Ref.
\cite{travers2014exponential} and references therein. Then, $\hmu(L)-\hmu \sim
Ke^{-\lambda L}$ in the large $L$ limit. From Eq.~(\ref{eq:EEL}), we see that
the convergence rate $\lambda$ also implies an exponential rate of decay for
$\EE-\EE(L) \sim K' e^{-\lambda L}$. Additionally, according to the asymptotic
equipartition property \cite{Cove06a}, when $L$ is large, $|\FeatAlphabet|\sim
e^{h_0 L}$ and $\H(L) \approx \hmu L$, where $h_0$ is the \emph{topological
entropy} and $\hmu$ the entropy rate, both in nats.

In sum, this first set of predictive features---effectively, the construction
of order-$L$ Markov models---yields an algebraic tradeoff between the size of
the feature set $\FeatAlphabet$ and the predictive distortion $d$:
\begin{align}
|\FeatAlphabet| \sim \left(\frac{1}{d}\right)^{h_0/\lambda},
\label{eq:OrderL_FScaling}
\end{align}
and a logarithmic tradeoff between the entropy of the features and distortion:
\begin{align}
H[\Rep] \sim \frac{\hmu}{\lambda} \log \left(\frac{1}{d}\right)
  ~.
\label{eq:OrderL_RepScaling}
\end{align}
In principle, $\lambda$ can be arbitrarily small, and so $h_0 / \lambda$ and
$\hmu / \lambda$ arbitrarily large, even for processes generated by
finite-state HMMs. This can be true even for finite unifilar HMMs (\eMs). To
see this, let $W$ be the transition matrix of a process' mixed-state
presentation; defined shortly. From Ref. \cite{Ara14a}, when $W$'s spectral gap
$\gamma$ is small, we have $\EE-\EE(L) \sim (1-\gamma)^L$, so that $\lambda =
\log\frac{1}{1-\gamma}$ can be quite small.

In short, when the process in question has only a finite number of causal
states, $|\FeatAlphabet|$ optimally saturates at $|\St|$ and $H[\Rep]$
optimally saturates at $\Cmu$, but from Eqs.~(\ref{eq:OrderL_FScaling}) and
(\ref{eq:OrderL_RepScaling}), the size of this Markov feature set can grow
without bound when we attempt to achieve zero predictive distortion.

\begin{figure*}
\centering
\includegraphics[height=0.3\textwidth]{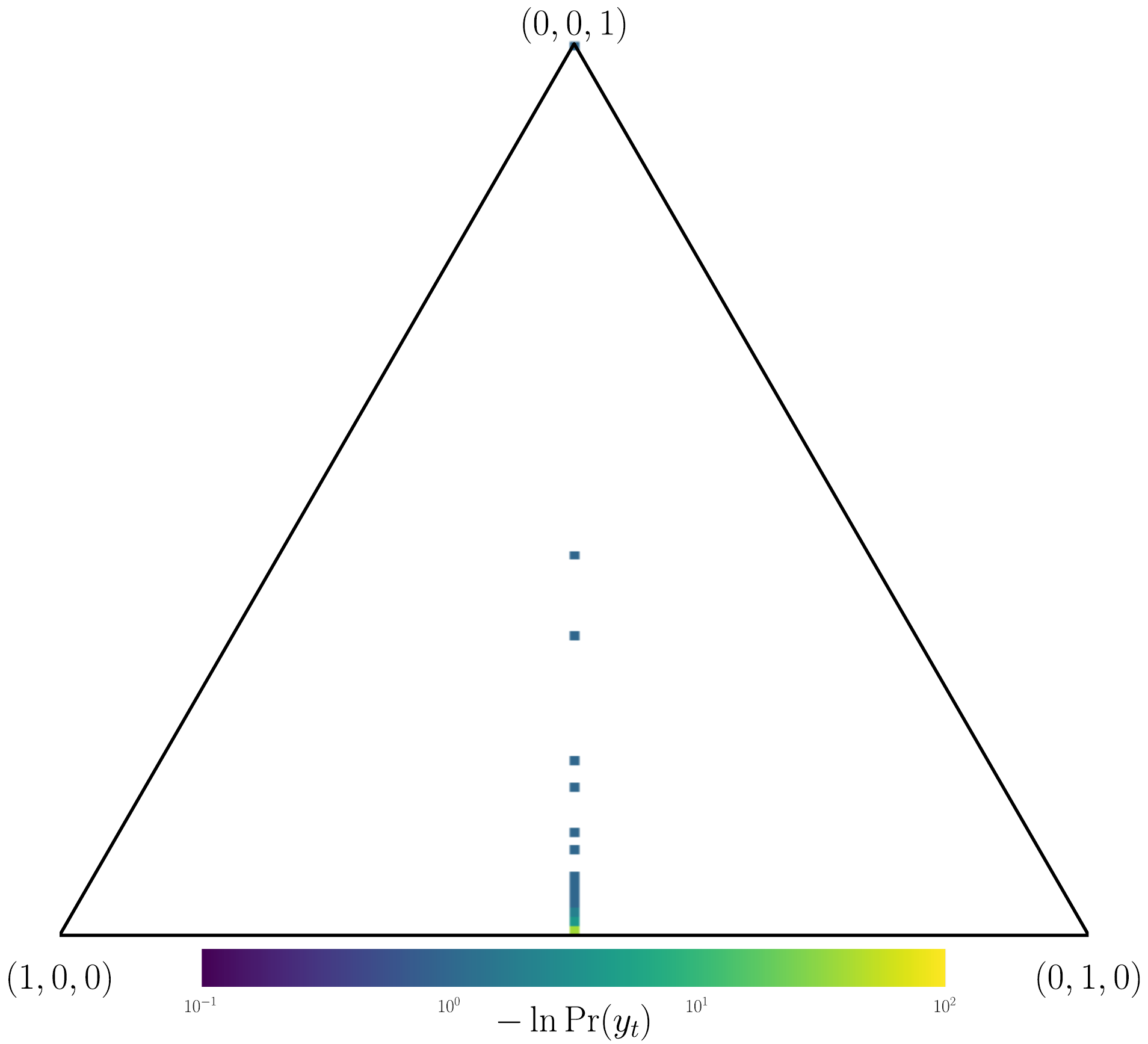}
\includegraphics[height=0.3\textwidth]{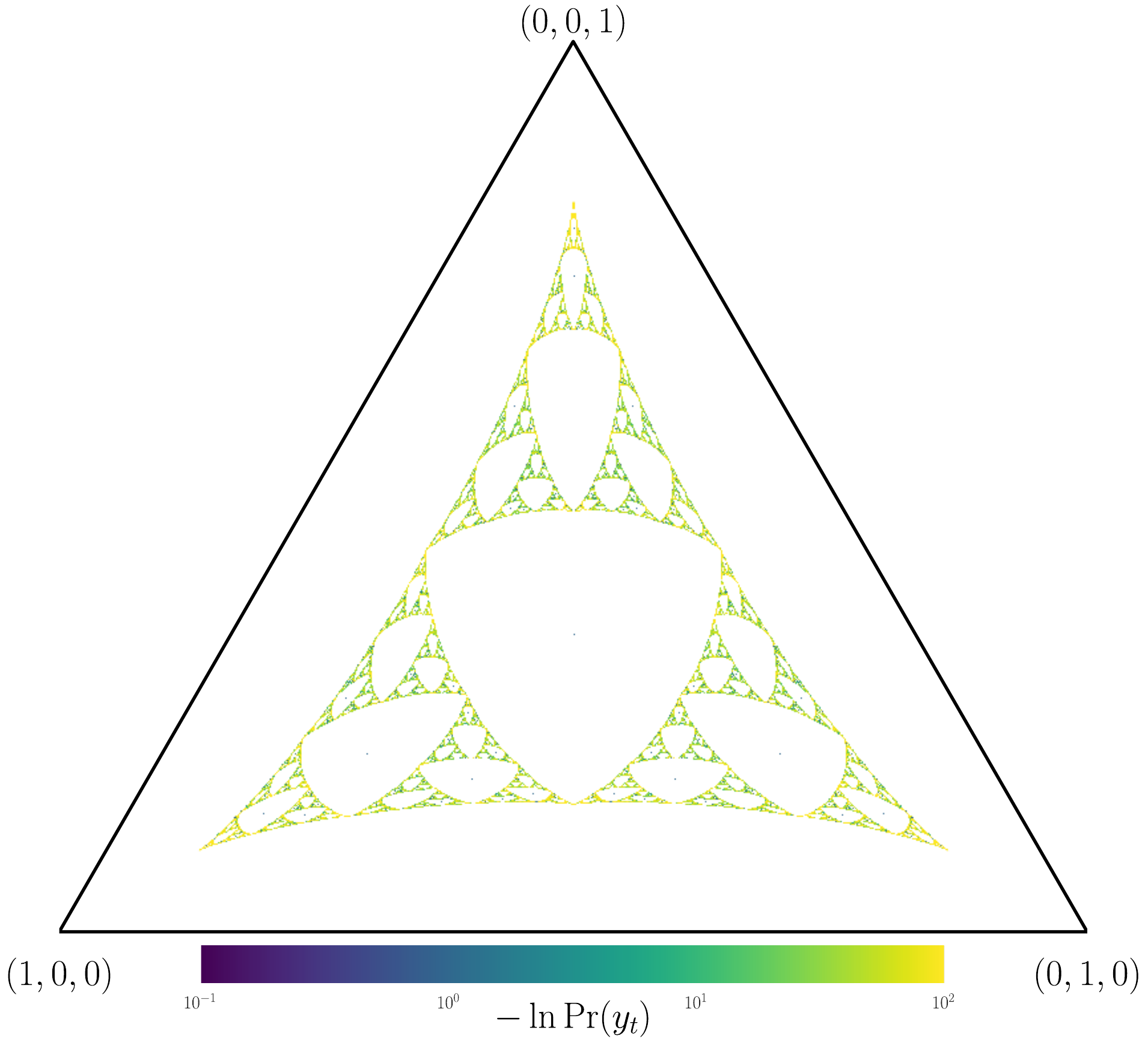}
\includegraphics[height=0.3\textwidth]{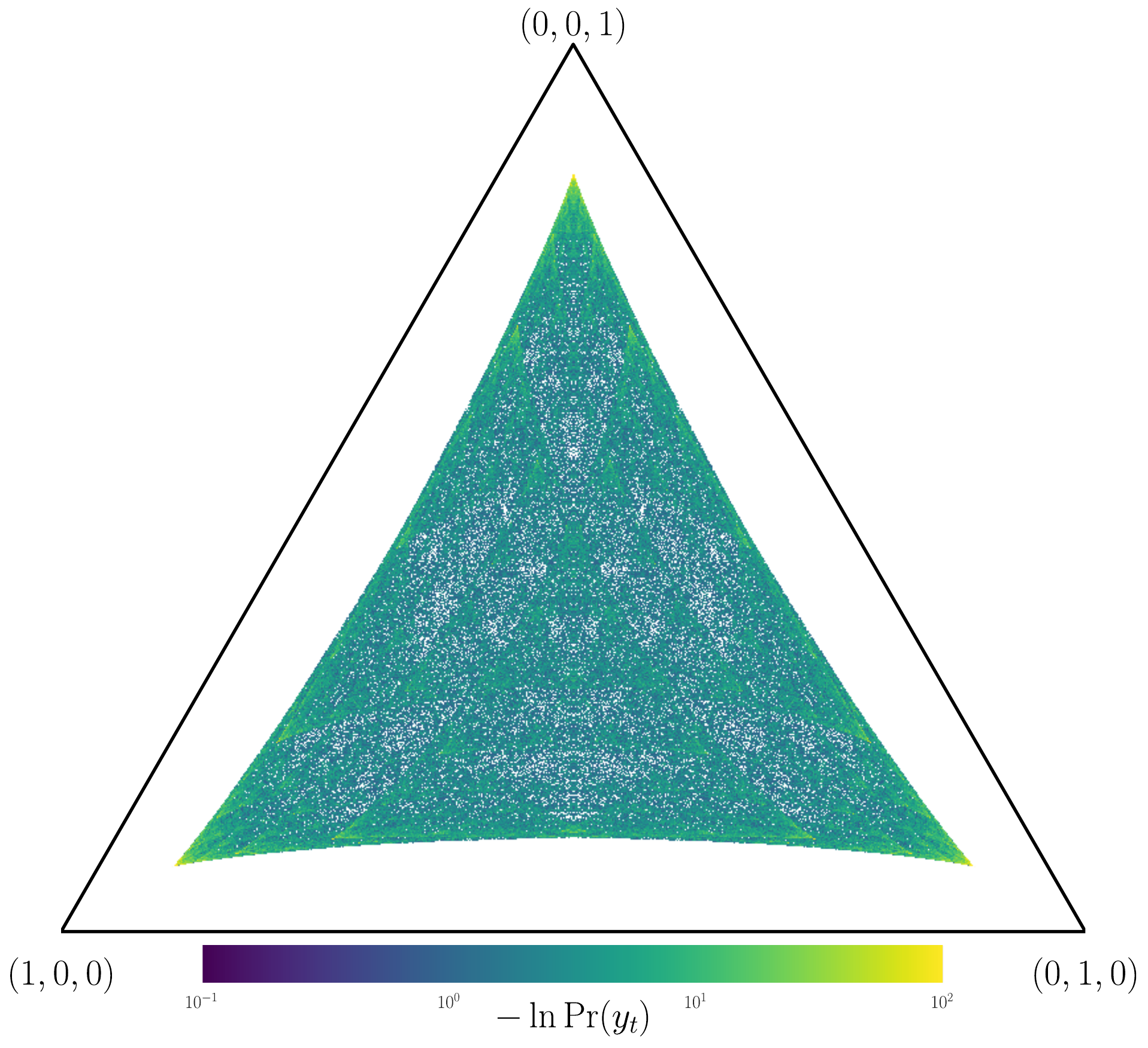}
\caption{Mixed state presentations $Y$ for three different generative models
	given in Supplementary Materials App.~\ref{app:EstimateDim}.
To visualize the mixed-state simplex,
	the diagrams plot iterates $y_t$ such that the left vertex corresponds to
	the pure state $\Pr(A,B,C) = (1,0,0)$ or HMM state $A$, the right to pure
	state $(0,1,0)$ or HMM state $B$, and the top to pure state $(0,0,1)$ or
	HMM state $C$.  Left plot has $55$ mixed states plotted in a $1 \times 100$
	bin histogram; middle plots $2,391,484$ mixed states in a $1000 \times 1000$
	bin histogram; and right plots $21,523,360$ mixed states in a $4000 \times
	4000$ bin histogram. Bin cell coloring is negative logarithm of the
	normalized bin counts. The box-counting dimensions are respectively:
	$\text{dim}_0(Y) \approx 0.3$ at left, $\text{dim}_0(Y) \approx 1.8$ in the
	middle, and $\text{dim}_0(Y) \approx 1.9$ at right. Box-counting dimension
	calculated by estimating the slope of $\log (1/\epsilon)$ versus $\log
	N_{\epsilon}$ as described in Supplementary Materials App.~\ref{app:EstimateDim}.
	}
\label{fig:MSPs}
\end{figure*}

\paragraph*{Mixed-State Features}
A different predictive feature set comes from coarse-graining the mixed-state
simplex. As first described by Blackwell \cite{Blac57b}, mixed states $Y$ are
probability distributions $\Prob(\Gen_0|\MS_{-L:0}=\ms_{-L:0})$ over the
internal states $\Gen$ of a generative model given the generated sequences.
Transient mixed states are those at finite $L$, while recurrent mixed
states are those remaining with positive probability in the limit that $L \to
\infty$. Recurrent mixed states exactly correspond to causal states $\St$
\cite{Crut08b}. When $\Cmu$ diverges, recurrent mixed states often lay on a
Cantor set in the simplex; see Fig. \ref{fig:MSPs}. In this circumstance, one
examines the various dimensions that describe the scaling in such sets. Here,
for reasons that will become clear, we use the box-counting $\text{dim}_0(Y)$
and information dimensions $\text{dim}_1(Y)$ \cite{Pesi97a}.

More concretely, we partition the simplex into cubes of side length $\epsilon$,
and each nonempty cube is taken to be a predictive feature in our
representation. When there are only a finite number of causal states, then this
feature set consists of the causal states $\St$ for some nonzero $\epsilon$.
There is no such $\epsilon$ if, instead, the process has a countable infinity
of causal states. Sometimes, as for the finite $|\St|$ case, the
information dimension and perhaps even the box-counting dimension of the mixed
states will vanish. When there is an uncountable infinity of causal states and
$\epsilon$ is sufficiently small, the corresponding number of features scales
as:
\begin{align*}
|\FeatAlphabet| \sim \left(\frac{1}{\epsilon}\right)^{\text{dim}_0(Y)}
  ~,
\end{align*}
where $\text{dim}_0(Y)$ denotes the box-counting dimension \footnote{We assume
here that the lower box-counting dimension and upper box-counting dimension are
equivalent.} and the coding cost scales as:
\begin{align*}
\H[\Rep] \sim \text{dim}_{1}(Y)\log\frac{1}{\epsilon}
  ~,
\end{align*}
where $\text{dim}_1(Y)$ is the information dimension. These dimensions can be
approximated numerically; see Fig.~\ref{fig:MSPs} and
Supplementary Materials App.~\ref{app:EstimateDim}.

For processes generated by infinite \eMs, finding the better feature set
requires analyzing the scaling of predictive distortion with $\epsilon$.
Supplementary Materials App. \ref{app:MSPbound}
shows that $d(\Rep)$ at coarse-graining $\epsilon$
scales at most as $\epsilon$ in the limit of asymptotically small $\epsilon$.
Our upper bound relies on the fact that two nearby mixed states have similar
future morphs, i.e., they have similar conditional probability distributions
over future trajectories given one's mixed state.  From this, we conclude that:
\begin{align}
|\FeatAlphabet| \sim \left(\frac{1}{d}\right)^{\text{dim}_0(Y)}
\label{eq:MSP_NumFeatures}
\end{align}
and
\begin{align}
\H[\Rep] \sim \text{dim}_1(Y)\log \frac{1}{d}
  ~.
\label{eq:MSP_CodeCost}
\end{align}
In general, both $\text{dim}_0(Y)$ and $\text{dim}_{1}(Y)$ are bounded above by $|\Gen|$, the number of states in the minimal generative model.

\paragraph*{Resource-Prediction Tradeoff}\label{sec:ResultsC}
Finally, putting Eqs.~(\ref{eq:OrderL_FScaling})-(\ref{eq:MSP_CodeCost})
together, we find that for a process with an uncountably infinite \eM, the
necessary number of predictive features $|\mathcal{F}^*|$ scales no faster
than:
\begin{align}
|\mathcal{F}^*| \lesssim \left(\frac{1}{d}\right)^{\min (h_0/\lambda,\text{dim}_0(Y))}
\label{eq:Fmin}
\end{align}
and the requisite coding cost $\H[\Rep^*]$ scales no faster than:
\begin{align}
\H[\Rep^*] \lesssim \min (\hmu/\lambda,\text{dim}_1(Y)) \log \frac{1}{d}
  ~,\label{eq:Hmin}
\end{align}
where $d$ is predictive distortion.

These bounds have a practical interpretation.  From finite data, one can only
justify inferring finite-state minimal maximally predictive models. Indeed, the
criteria of Ref. \cite{still2004many} applied as in Ref. \cite{Stil07b}
suggests that the maximum number of inferred states should not yield predictive
distortions below the noise in our estimate of predictive distortion $d$ from
$T$ data points. This noise scales as $\sim 1 / \sqrt{T}$, since from $T$ data
points, we have approximately $T$ separate measurements of predictive
distortion. This, in turn, sets upper bounds on $|\mathcal{F}^*| \lesssim
T^{\frac{1}{2}\min (h_0/\lambda,\text{dim}_0(Y))}$ and $\H[\Rep^*]\lesssim \min
(\hmu/\lambda,\text{dim}_1(Y)) \log T$ when the process in question has an
uncountable infinity of causal states.

%The bounds presented here give a sense of how many states might be
%inferred according to the clustering criteria of Ref. \cite{still2004many}
%applied to the mixed-state simplex distributions, paralleling that done in
%Ref. \cite{Still10a} for processes generated by finite-state \eMs.
%The standard deviation of predictive distortion, assuming that we are studying a process with finite excess entropy, will tend to
%decrease as the inverse square root of the amount of data given \alert{[CITE]}.
%As such, we might expect to see the number of states in the
%inferred minimal maximally predictive model scaling algebraically, with a coefficient upper-bounded by twice the box-counting dimension of the mixed state space.

We can test this prediction directly using the Bayesian Structural Inference
(BSI) algorithm \cite{Stre13a} applied to the processes described in
Fig.~\ref{fig:MSPs}.  The major difficulty in employing BSI is that one must
specify a list of \eM\ topologies to search over. Since the number of such
topologies grows super-exponentially with the number of states \cite{John10a},
experimentally probing the scaling behavior of the number of inferred states
with the amount of available data is, with nothing else said, impossible.

\begin{figure}
\centering
\includegraphics[width=0.45\textwidth]{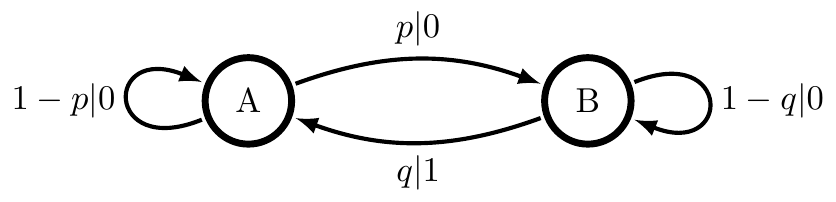}
\includegraphics[width=0.45\textwidth]{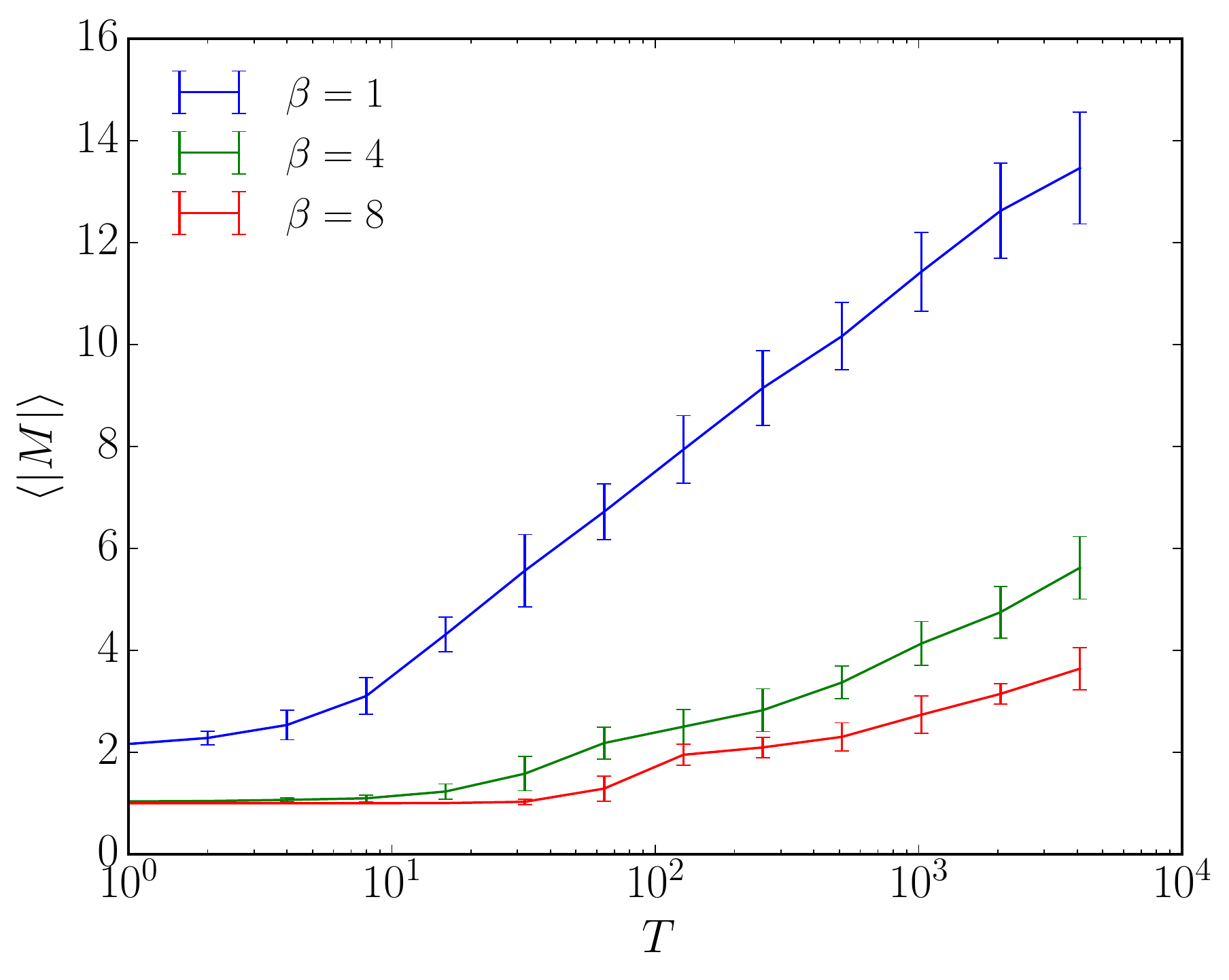}
\caption{Model-size scaling for the parametrized Simple Nonunifilar Source
	(SNS). (Top) Generative HMM for the SNS. (Bottom) Scaling of $\langle
	|M|\rangle = \sum_{M} |M| P(M|x_{0:T})$ with $T$, with the posterior
	$P(M|x_{0:T})$ calculated using formulae from BSI \cite{Stre13a} with
	$\alpha=1$ and the set of model topologies being the eventually Poisson
	\eMs~described in Ref. \cite{Marz14b}. Data generated from the SNS
	with $p=q=\frac{1}{2}$.  Linear regression reveals a slope of $\approx 1$
	for $\log\log T$ vs. $\log \langle |M|\rangle$, confirming the expected
	$N_T \approxprop\log T$, where the proportionality constants depend on
	$\beta$.
	}
\label{fig:SNS_Scaling}
\end{figure}

However, ``expert knowledge'' can cull the number of \eM\ topologies that one
should search over. In this spirit, we focus on the Simple Nonunifilar Source
(SNS), since \eMs\ for renewal processes have been characterized in detail
\cite{Marz14b,Marz15a,Marz14e,Marz17a}. The SNS's generative HMM is given in
Fig.~\ref{fig:SNS_Scaling}(left). This process has a mixed-state presentation
similar to that of ``nond'' in Fig.~\ref{fig:MSPs}(left).  We choose to only
search over the \eM\ topologies that correspond to the class of eventually
Poisson renewal processes.

We must first revisit the upper bound in Eq.~(\ref{eq:Fmin}), as the SNS has a
countable (not uncountable) infinity of causal states.  When a process' \eM\ is
countable instead of uncountable, then we can improve upon Eq.~(\ref{eq:Fmin}).
However, the magnitude of improvement is process dependent and not easily
characterized in general for two main reasons. First, predictive distortion can
decrease faster than $\epsilon$ for processes generated by countably infinite
\eMs\ since there might only be one mixed state in an $\epsilon$ hypercube.
(See the lefthand factor in Supplementary Materials Eq.~(\ref{eq:d_L_ubound}),
$\sum_{r\in\Rep^{(\epsilon)}:\H[Y|\Rep^{(\epsilon)}=r]=0} \pi(r)$. We are
guaranteed that $\lim_{\epsilon\rightarrow 0}
\sum_{r\in\Rep^{(\epsilon)}:\H[Y|\Rep^{(\epsilon)}=r]=0} \pi(r)=0$, but the
rate of convergence to zero is highly process dependent.) Second, the scaling
relation between $N_{\epsilon}$ and $1/\epsilon$ may be subpower law.

Given a specific process, though, with a countably infinite \eM---here, the
SNS---we can derive expected scaling relations. See Supplementary Materials
App. \ref{app:SNSExample}.
We argue that, roughly speaking, we should expect
predictive distortion to decay exponentially with $N_{\epsilon}$, so that
$d(\Rep_{\epsilon}) ~\approxprop~ \lambda^{N_{\epsilon}}$ for some $\lambda<1$.
As we expect, since our uncertainty in $d$ scales as $\sim 1 / \sqrt{T}$, where
$T$ is the amount of data, we expect that the number of inferred states $N_T$
should scale as $\approxprop~\log T$.

%We find that, when $p\neq q$, $N_{\epsilon} \sim \left(\log\log\frac{1}{d}\right)^2$, and when $p=q$, $N_{\epsilon}\sim\sqrt{\log\frac{1}{d}}$.  As we expect our uncertainty in $d$ to scale $\sim\frac{1}{\sqrt{T}}$ where $T$ is the amount of data, we expect the number of inferred states $\sim \left(\log\log T\right)^2$ when $p\neq q$ and $\sim \sqrt{\log T}$ when $p=q$.

These scaling relationships are confirmed by BSI applied to data generated from
the SNS, where the set of \eM\ topologies selected from are the eventually
Poisson \eM\ topologies characterized by Ref. \cite{Marz14b}. Given a set of
machines $\mathcal{M}$ and data $x_{0:T}$, BSI returns an easily-calculable
posterior $\Pr(M|x_{0:T})$ for $M\in\mathcal{M}$ with (at least) two
hyperparameters: (i) the concentration parameter for our Dirichlet prior on the
transition probabilities $\alpha$ and (ii) our prior on the likelihood of a
model $M$ with number of states $|M|$, taken to be proportional to $e^{-\beta
|M|}$ for a user-specified $\beta$. From this posterior, we calculate an
average model size $\langle |M|\rangle(T) = \sum_{M\in\mathcal{M}} |M|
\Pr(M|x_{0:T})$ as a function of $T$ for multiple data strings $x_{0:T}$.
% The BSI algorithm has (at least) two hyperparameters: $\alpha$, which specifies a prior on the distribution of transition probabilities and $\beta$, which gives a complexity-controlling prior $\propto e^{-\beta|M|}$ on the size of the true \eM.
The result is that we find the exact scaling of $\langle |M|\rangle(T)$
with $T$ is proportional to $\log T$ in the large $T$ limit, where the proportionality constant and
the initial value depends on hyperparameters $\alpha$ and $\beta$.

\paragraph*{Conclusion}
We now better understand the tradeoff between memory and prediction in
processes generated by a finite-state Hidden Markov model with infinite
statistical complexity. Importantly and perhaps still underappreciated, these
processes are more ``typical'' than those with finite statistical complexity
and Gaussian processes, which have received more attention in related literature
\cite{creutzig2009past, Stil07b}.

We proposed a new method for obtaining predictive features---coarse-graining
mixed states---and used this feature set to bound the number of features and
their coding cost in the limit of small predictive distortion. These bounds
were compared to those obtained from the more familiar and widely used (Markov)
predictive feature set---that of memorizing all pasts of a fixed length. The
general results here suggest that the new feature set can outperform the
standard set in many situations.

Practically speaking, the new bounds presented have three potential uses.
First, they give weight to Ref. \cite{Crut92c}'s suggestion to replace the
statistical complexity with the information dimension of mixed-state space as a
complexity measure when statistical complexity diverges. Second, they suggest a
route to an improved, process-dependent tradeoff between complexity and
precision for estimating the entropy rate of processes generated by nonunifilar
HMMs \cite{ordentlich2011bounds}. (Admittedly, here space kept us from
addressing how to estimate the probability distribution over $\epsilon$-boxes.)

Finally, and perhaps most importantly, our upper bounds provide a first attempt
to calculate the expected scaling of inferred model size with available data.
It is commonly accepted that inferring time-series models from finite data
typically has two components---parameter estimation and model
selection---though these two can be done simultaneously. Our focus above was on
model selection, as we monitored how model size increased with the amount of
available data. One can view the results as an effort to characterize the
posterior distribution of \eM\ topologies given data or as the growth rate of
the optimum model cost \cite{Riss84a} in the asymptotic limit. Posterior
distributions of estimated parameters (transition probabilities) are almost
always asymptotically normal, with standard deviation decreasing as the square
root of the amount of available data \cite{walker1969asymptotic,
heyde1979asymptotic, sweeting1992asymptotic}. However, asymptotic normality
does not typically hold for this posterior distribution, since using finite
\eM\ topologies almost always implies out-of-class modeling. Rather, we showed
that the particular way in which the mode of this posterior distribution
increases with data typically depends on a gross process statistic---the
box-counting dimension of the mixed-state presentation.

Stepping back, we only tackled the lower parts of the \emph{infinitary} process
hierarchy identified in Ref. \cite{Crut15a}. In particular, ``complex''
processes in the sense of Ref. \cite{Bial00a} have different
resource-prediction tradeoffs than analyzed here, since processes generated by
finite-state HMMs (as assumed here) cannot produce processes with infinite
excess entropy. To do so, at the very least, predictive distortion must be more
carefully defined. We conjecture that success will be achieved by instead
focusing on a one-step predictive distortion, equivalent to a self-information
loss function, as is typically done \cite{merhav1998universal, Riss84a}.
Luckily, the derivation in Supplementary Materials App.~\ref{app:MSPbound}
easily extends to this case, simultaneously suggesting improvements to related
entropy rate-approximating algorithms.

We hope these introductory results inspire future study of resource-prediction
tradeoffs for infinitary processes.

% \acknowledgments

The authors thank Santa Fe Institute for its hospitality during visits and
A. Boyd, C. Hillar, and D. Upper for useful discussions. JPC is an SFI
External Faculty member. This material is based upon work supported by, or in
part by, the U.S. Army Research Laboratory and the U. S. Army Research Office
under contracts W911NF-13-1-0390 and W911NF-12-1-0288. S.E.M. was funded by a
National Science Foundation Graduate Student Research Fellowship, a U.C.
Berkeley Chancellor's Fellowship, and the MIT Physics of Living Systems
Fellowship.

%=================================================================
% References:
%=================================================================
% Use the following option to include external BibTeX files:
\bibliography{chaos}

\onecolumngrid
\appendix
\clearpage

\input{supp_final}

\end{document}

%% file: cmechabbrev.tex
\newcommand{\eM}     {\mbox{$\epsilon$-machine}}
\newcommand{\eMs}    {\mbox{$\epsilon$-machines}}

\newcommand{\Process}{\mathcal{P}}
\newcommand{\Rep}{\mathcal{R}}

\newcommand{\Gen}{\mathcal{G}}

\newcommand{\MeasAlphabet}  {\mathcal{A}}
\newcommand{\MeasSymbol}   { {X} }
\newcommand{\meassymbol}   { {x} }
\newcommand{\MeasSymbols}[2]{ \MeasSymbol_{#1:#2} }
\newcommand{\meassymbols}[2] { \meassymbol_{#1:#2} }

\newcommand{\Past} { \MeasSymbols{}{0} }
\newcommand{\past} { \meassymbols{}{0} }

\newcommand{\Future} { \MeasSymbols{0}{} }

\newcommand{\FutureL}     { \smash{\overrightarrow{\MeasSymbol}{}^L} }

\newcommand{\CausalState}   { \mathcal{S} }

\newcommand{\causalstate}   { \sigma }
\newcommand{\CausalStateSet}    { \boldsymbol{\CausalState} }
\newcommand{\AlternateState}    { \mathcal{R} }

\newcommand{\AlternateStateSet} { \boldsymbol{\AlternateState} }

\newcommand{\Prob}      {\Pr} % use standard command

\newcommand{\Cmu}       {C_\mu}
\newcommand{\hmu}       {h_\mu}
\newcommand{\EE}        {{\bf E}}

\newcommand{\ProcessAlphabet}   {\MeasAlphabet}

\newcommand{\forward}{+}
\newcommand{\reverse}{-}
\newcommand{\forwardreverse}{\pm} % \pm

\newcommand{\FutureCausalState} { {\CausalState}^{\forward} }

\newcommand{\PastCausalState}   { {\CausalState}^{\reverse} }

\newcommand{\lastindex}[2]{
  \edef\tempa{0}
  \edef\tempb{#2}
  \ifx\tempa\tempb
    \edef\tempc{#1}
  \else
    \edef\tempa{0}
    \edef\tempb{#1}
    \ifx\tempa\tempb
      \edef\tempc{#2}
    \else
      \edef\tempc{#1+#2}
    \fi
  \fi
  \tempc
}

\newcommand{\I}{\mathbf{I}}

\newcommand{\CSjoint}[1][,]{
   \edef\tempa{:}
   \edef\tempb{#1}
   \ifx\tempa\tempb
      \ensuremath{\FutureCausalState\!#1\PastCausalState}
   \else
      \ensuremath{\FutureCausalState#1\PastCausalState}
   \fi
}

\newif\ifpm
\edef\tempa{\forwardreverse}
\edef\tempb{\pm}
\ifx\tempa\tempb
   \pmfalse
\else
   \pmtrue
\fi

%% file: supp_final.tex
\begin{center}
\large{Supplementary Materials}\\
\normalsize
for\\
\emph{Nearly Maximally Predictive Features and Their Dimensions}\\
Sarah E. Marzen and James P. Crutchfield
\end{center}

\section{Estimating fractal dimensions}
\label{app:EstimateDim}

We start by describing the generative models with mixed-state presentations
shown in Fig.~\ref{fig:MSPs}. Figure~\ref{fig:MSPs}(left) has labeled transition matrices of:
\begin{align*}
T^{(0)} = \begin{pmatrix} 0 & 0 & 0 \\ 0 & 0 & 0 \\ 1 & 0 & 0 \end{pmatrix}
\text{ and }
T^{(1)} = \begin{pmatrix} 0 & 1/2 & 1/3 \\ 0 & 1/2 & 1/3 \\ 0 & 0 & 1/3 \end{pmatrix}
  ~.
\end{align*}
Figure~\ref{fig:MSPs}(middle) and Fig.~\ref{fig:MSPs}(right) have labeled transition matrices parametrized by $x$ and $\alpha$:
\begin{align*}
T^{(0)} = \begin{pmatrix} \alpha y & \beta x & \beta x \\ \alpha x & \beta y & \beta x \\ \alpha x & \beta x & \beta y \end{pmatrix}
  ~, ~
T^{(1)} = \begin{pmatrix} \beta y & \alpha x & \beta x \\ \beta x & \alpha y & \beta x \\ \beta x & \alpha x & \beta y \end{pmatrix}
  ~, ~ \text{and }
T^{(2)} = \begin{pmatrix} \beta y & \beta x & \alpha x \\ \beta x & \beta y & \alpha x \\ \beta x & \beta x & \alpha y \end{pmatrix}
  ~,
\end{align*}
with $\beta = \frac{1-\alpha}{2}$ and $y=1-2x$. Figure~\ref{fig:MSPs}(middle) corresponds to $x=0.15$ and $\alpha = 0.6$, while Fig.~\ref{fig:MSPs}(right) corresponds to $x=0.05$ and $\alpha = 0.6$.
 
The probabilities of emitting a $0$, $1$, and $2$ given the current state are,
respectively:
\begin{align*}
p(\ms|A) & =
  \begin{cases} \alpha & \ms=0 \\ \frac{1-\alpha}{2} & \ms=1,2 \end{cases}
  ~, \\
p(\ms|B) & =
  \begin{cases} \alpha & \ms=1 \\ \frac{1-\alpha}{2} & \ms=0,2 \end{cases}
  ~,~\text{and} \\
p(\ms|C) & =
  \begin{cases} \alpha & \ms=2 \\ \frac{1-\alpha}{2} & \ms=0,1 \end{cases}
  ~.
\end{align*}

\begin{figure}[h]
\centering
\includegraphics[width=0.45\textwidth]{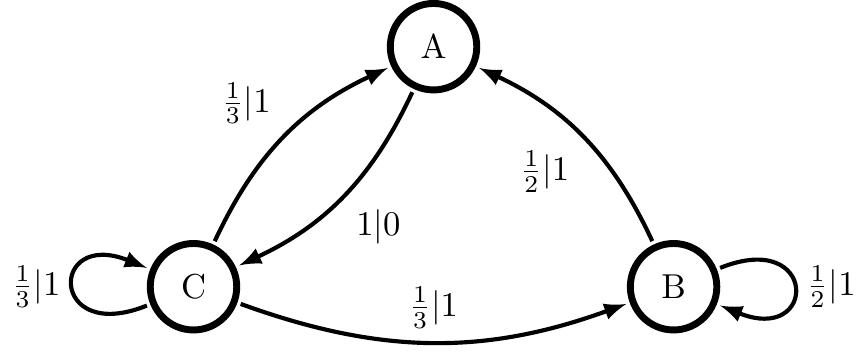}
\includegraphics[width=0.45\textwidth]{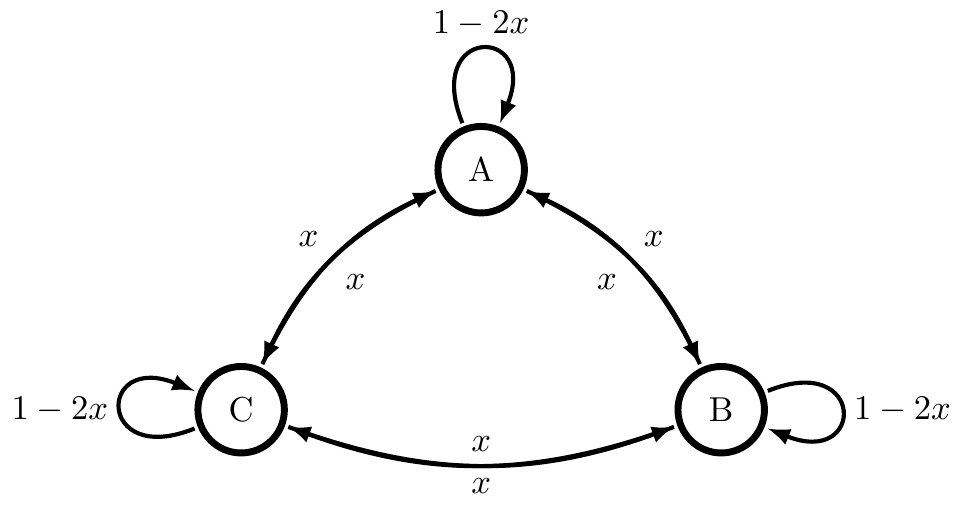}
\caption{(Left) Minimal generative model for the ``nond'' process whose
	mixed-state presentation is shown in Fig.~\ref{fig:MSPs}(left). (Right)
	Parametrized minimal generative model for the ``mess3'' process. For
	Fig.~\ref{fig:MSPs}(middle), $x=0.15$ and $\alpha = 0.6$ and for
	Fig.~\ref{fig:MSPs}(right), $x=0.05$ and $\alpha=0.6$.
	}
\end{figure}
 
The box-counting dimension is approximated by forming the mixed-state
presentation from the generative model as described by Ref. \cite{Crut08b} down
to the desired tree depth.  For Fig.~\ref{fig:MSPs}(left), the depth was
sufficient to accrue $10,000$ mixed states; for Fig.~\ref{fig:MSPs}(middle) and
Fig.~\ref{fig:MSPs}(right), the depth was sufficient to accrue $30,000$ mixed
states.  Box-counting dimension was estimated by partitioning the simplex into
boxes of side length $\epsilon = 1/n$ for $n \in \{1,...,300\}$, calculating
the number $N_{\epsilon}$ of nonempty boxes, and using linear regression on
$\log N_{\epsilon}$ versus $\log \frac{1}{\epsilon}$ to find the slope.
Information dimension could be estimated as well by estimating probabilities of
words that lead to each mixed state.

\section{Predictive distortion scaling for coarse-grained mixed states}
\label{app:MSPbound}

\newcommand\norm[1]{\left\lVert#1\right\rVert}

Recall that the second feature set (alphabet $\FeatAlphabet$, random variable $\Rep$) is constructed by partitioning the simplex
into boxes of side-length $\epsilon$. This section finds the scaling of $d(\Rep)$ with $\epsilon$ in the limit of asymptotically small $\epsilon$.

For reasons that become clear shortly, we define:
\begin{align}
d_L(\Rep) & = \I[\Past;\FutureL|\Rep] \\
          & = \H[\FutureL|\Rep] - \H[\FutureL|\Past]
  ~,
\label{eq:d_L}
\end{align}
where $\lim_{L\rightarrow\infty} d_L(\Rep) = d(\Rep)$.  Let $\pi(r)$ be the invariant probability distribution over $\epsilon$-boxes, and let $\pi(y|r)$ be the probability measure over mixed states in that $\epsilon$-box.  Then:
\begin{align}
d_L(\Rep) = \sum_r \pi(r)
  \left( \H[\FutureL|\Rep=r]-\int_{y} d\pi(y|r)  \H[\FutureL|\Gen\sim y]\right)
  ~,
\label{eq:d_L_2}
\end{align}
where:
\begin{align}
\Prob(\FutureL|\Rep=r) = \int_y d\pi(y|r) \Prob(\FutureL|\Gen\sim y)
  ~.
\end{align}
To place an upper bound on $d_L(\Rep)$ in terms of $\epsilon$, we note that for
any two mixed states $y$ and $y'$ in the same partition:
\begin{align*}
\norm{y-y'}_1 \leq \sqrt{|\Gen|}\epsilon
  ~,
\end{align*}
the length of the longest diagonal in a hypercube of dimension $|\Gen|$, by
construction. Hence:
\begin{align*}
\norm{\Prob(\FutureL|\Gen\sim y)-\Prob(\FutureL|\Gen\sim y')}_{TV}
  & = \norm{\sum_{i=1}^{|\Gen|} \Prob(\FutureL|\Gen=i) (y_i-y'_i)}_{TV} \\
  & \leq \sum_{i=1}^{|\Gen|} \norm{\Prob(\FutureL|\Gen=i)}_{TV} |y_i-y'_i| \\
  & = \sum_{i=1}^{|\Gen|} |y_i-y'_i|
  ~.
\end{align*}
From earlier, though, this is simply:
\begin{align*}
\norm{\Prob(\FutureL|\Gen\sim y)-\Prob(\FutureL|\Gen\sim y')}_{TV}
  & \leq \norm{y-y'}_1 \\
  & \leq \sqrt{|\Gen|} \epsilon
  ~.
\end{align*}
(Often the literature defines $\norm{\cdot}_{TV}$ as $1/2$ of the quantity used
here.) From this, we can similarly conclude that:
\begin{align*}
\norm{\Prob(\FutureL|\Rep=r) - \Prob(\FutureL|\Gen\sim y')}_{TV}
  & = \norm{ \int_y d\pi(y|r) \Prob(\FutureL|\Gen\sim y) -
  \Prob(\FutureL|\Gen\sim y') }_{TV} \\
  & \leq \int_y d\pi(y|r)
    \norm{\Prob(\FutureL|\Gen\sim y)-\Prob(\FutureL|\Gen\sim y')}_{TV} \\
  & \leq \int_y d\pi(y|r) \sqrt{|\Gen|} \epsilon = \sqrt{|\Gen|} \epsilon
\end{align*}
for any mixed state $y'$ in partition $r$. Reference \cite{ho2010interplay}
gives us an upper bound on differences in entropy in terms of the total
variation, which here implies that:
\begin{align*}
\left|\H[\FutureL|\Rep=r]-\H[\FutureL|\Gen\sim y'] \right|
  \leq \H_b \left(\frac{\sqrt{|\Gen|}\epsilon}{2}\right)
  + \frac{\epsilon L \log_2 |\MeasAlphabet|}{2}
  ~.
\end{align*}
This, in turn, gives:
\begin{align*}
\left|H[\FutureL|\Rep=r]-\int_y d\pi(y|r) H[\FutureL|\Gen\sim y] \right|
  & \leq \int_y d\pi(y|r)
    \left| H[\FutureL|\Rep=r]-H[\FutureL|\Gen\sim y] \right| \\
  & \leq H_b\left(\frac{\sqrt{|\Gen|}\epsilon}{2}\right)+\frac{\epsilon L \log_2 |\MeasAlphabet|}{2}
  ~.
\end{align*}
With that having been said, if there is only one mixed state in some
$\epsilon$-cube $r$, the quantity above vanishes:
\begin{align*}
\left| \H[\FutureL|\Rep=r]-\int_y d\pi(y|r) H[\FutureL|\Gen\sim y] \right| = 0
  ~.
\end{align*}
Denote the probability of a nonempty $\epsilon$-cube having only one mixed
state in it as $\sum_{r\in\FeatAlphabet:H[Y|\Rep=r]=0} \pi(r)$. Then substitution into
Eq.~(\ref{eq:d_L_2}) gives:
\begin{align}
d_L(\Rep) \leq \left(\sum_{r\in\FeatAlphabet:H[Y|\Rep=r]=0} \pi(r) \right)\left(H_b\left(\frac{\sqrt{|\Gen|}\epsilon}{2}\right)+\frac{\epsilon L \log_2 |\MeasAlphabet|}{2}\right).
\label{eq:d_L_ubound}
\end{align}
Note that this implies the entropy rate can be approximated with error no greater than $\left(H_b\left(\frac{\sqrt{|\Gen|}\epsilon}{2}\right)+\frac{\epsilon \log_2 |\MeasAlphabet|}{2}\right)$ if one knows the distribution over $\epsilon$-boxes, $\pi(r)$.
The left factor $\sum_{r\in\FeatAlphabet:H[Y|\Rep=r]=0} \pi(r)$ tends to zero as the
partitions shrink for a process generated by a countable \eM\ and not so
otherwise.  For ease of presentation, we first study the case of uncountably
infinite \eMs, and then comment on the case of processes produced by countable
\eMs.

For clarity, we now introduce the notation $\Rep^{(\epsilon)}$ to indicate the
predictive features obtained from an $\epsilon$-partition of the mixed-state
simplex. From Eq.~(\ref{eq:d_L_ubound}), we can readily conclude that for any finite $L$:
\begin{align*}
\lim_{\epsilon\rightarrow 0} \frac{d_L(\Rep^{(\epsilon)})}{\epsilon^{\gamma}} = 0
  ~,
\end{align*}
when $\gamma<1$. We would like to extend this conclusion to
$d(\Rep^{(\epsilon)})$, but inspection of Eq.~(\ref{eq:d_L_ubound}) suggests
that concluding anything similar for $d(\Rep^{(\epsilon)})$ requires care. In
particular, we would like to show that $\lim_{\epsilon\rightarrow 0}
d(\Rep^{(\epsilon)}) / \epsilon^{\gamma} = 0$ for any $\gamma<1$.  Recall
that $d(\Rep) = \lim_{L\rightarrow\infty} d_L(\Rep)$. \textit{If} we can
exchange the limits $\lim_{\epsilon\rightarrow 0}$ and
$\lim_{L\rightarrow\infty}$, then:
\begin{align*}
\lim_{\epsilon\rightarrow 0} \frac{d(\Rep^{(\epsilon)})}{\epsilon^{\gamma}}
  & = \lim_{\epsilon\rightarrow 0} \lim_{L\rightarrow \infty}
  \frac{d_L(\Rep^{(\epsilon)})}{\epsilon^{\gamma}} \\
  & = \lim_{L\rightarrow\infty} \lim_{\epsilon\rightarrow 0}
  \frac{d_L(\Rep^{(\epsilon)})}{\epsilon^{\gamma}} \\
  & = 0
  ~.
\end{align*}
To justify exchanging limits, we appeal to the Moore-Osgood Theorem
\cite{tao2011introduction}. Consider any monotone decreasing sequence
$\{\epsilon_i\}_{i=1}^{\infty}$ of $\epsilon$, such that $a_{i,L} :=
d_L(\Rep^{(\epsilon_i)}) / \epsilon_i^{\gamma}$ is a doubly-infinite sequence.
When $\gamma<1$, we have that $\lim_{i\rightarrow\infty} a_{i,L} = 0$.  We
provide a plausibility argument for uniform convergence of $a_{i,L}$ to $0$.
Recall from Eq.~(\ref{eq:d_L_ubound}) that:
\begin{align*}
|a_{i,L}-0|=a_{i,L}\leq \frac{1}{\epsilon_i^{\gamma}}\left(H_b(\frac{\sqrt{|\Gen|}\epsilon_i}{2}) + \frac{\epsilon_i L \log_2|\MeasAlphabet|}{2}\right)
  ~.
\end{align*}
And, since $H_b(x) \leq -x\log_2 x + x$, this expands to:
\begin{align*}
a_{i,L}\leq \frac{\sqrt{|\Gen|}\epsilon_i^{1-\gamma}}{2}\log_2 \left(1/\epsilon_i\right) + \left(\frac{\sqrt{|\Gen|}}{2}\log_2 \frac{2}{\sqrt{|\Gen|}} + \frac{\sqrt{|\Gen|}}{2}+\frac{L\log_2|\MeasAlphabet|}{2}\right)\epsilon_i^{1-\gamma}
  ~.
\end{align*}
At small enough $\epsilon_i$, the first term dominates, implying that $a_{i,L}
\leq \sqrt{|\Gen|}\epsilon_i^{1-\gamma}\log_2 \frac{1}{\epsilon_i}$. By making
$i$ sufficiently large (i.e., by making $\epsilon_i$ sufficiently small) we can
make this upper bound as small as desired. Finally, the Data Processing
Inequality implies that $\I[\Past;\FutureL|\Rep]$ is monotone increasing in $L$
and upper bounded by $\EE<\log_2|\Gen|<\infty$ \footnote{This upper bound on
$\EE$ was noted in Ref. \cite{lohrthesis}.}. And so, $a_{i,L}$ is also monotone
increasing in $L$ and upper-bounded by $\EE / \epsilon_i^{\gamma}$. Hence, the
monotone convergence theorem implies that $\lim_{L\rightarrow\infty} a_{i,L}$
exists for any $i$. Therefore, the conditions of the Moore-Osgood Theorem are
satisfied, and $\lim_{i\rightarrow\infty} \lim_{L\rightarrow\infty}
d_L(\Rep^{(\epsilon_i)}) / \epsilon_i^{\gamma} = 0$, so
$\lim_{\epsilon\rightarrow 0} d(\Rep^{(\epsilon)}) / \epsilon^{\gamma} = 0$ for
any $\gamma<1$ as desired. Loosely speaking, as in the main text, we say simply
that $d(\Rep^{(\epsilon)})$ scales as $\epsilon$ for small $\epsilon$.

\section{Countably infinite \eM\ example: the simple nonunifilar source}
\label{app:SNSExample}

\begin{figure}
\includegraphics[width=0.48\textwidth]{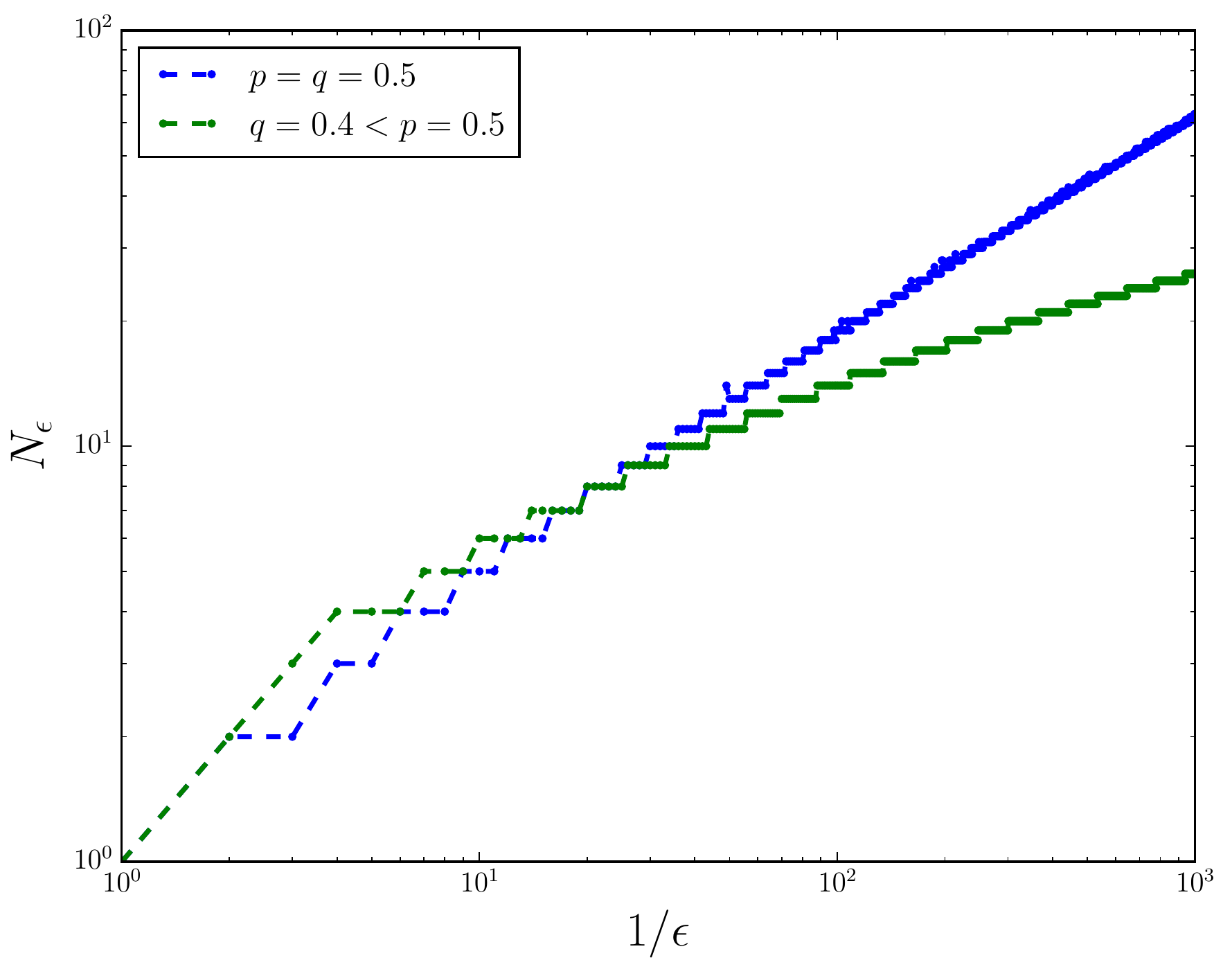}
\caption{Feature-set scaling of SNS mixed states: Number $N_{\epsilon}$ of
	nonempty $\epsilon$-cubes as a function of $1/\epsilon$ on a log-log plot.
	When $p=q$, the scaling relation appears to be power law: $N_{\epsilon}
	\approx (1/\epsilon)^{1/2}$. When $p\neq q$, $N_{\epsilon}$ scales more
	slowly than a power law with $1/\epsilon$, seemingly at a rate
	$N_{\epsilon}\sim \left(\log \frac{1}{\epsilon}\right)^2$.
	}
\label{fig:SNS}
\end{figure}

Consider the parametrized Simple Nonunifilar Source (SNS), represented by
Fig.~\ref{fig:SNS}. It is straightforward to show that the mixed states lie on
a line in the simplex $\{(v_n,1-v_n)\}_{n=0}^{\infty}$ with:
\begin{align*}
v_n = \frac{(1-p)^n (q-p)}{(1-p)^n q - (1-q)^n p}
  ~.
\end{align*}
If $p<q$, then $\lim_{n\rightarrow\infty} v_n = 1- p/q$; if $q<p$, then
$\lim_{n\rightarrow\infty} v_n = 0$.

It is also straightforward to show that $v_n$ converges exponentially fast to
its limit when $p\neq q$:
\begin{align*}
v_n - \lim_{n\rightarrow\infty} v_n \approx \begin{cases} -(1-\frac{p}{q})(\frac{p}{q}(\frac{1-q}{1-p})^n) & p<q \\ (1-\frac{q}{p}) \left(\frac{1-p}{1-q}\right)^{n} & q<p \end{cases}
  ~.
\end{align*}

However, when $p=q$, then $v_n = \frac{n}{n+(\frac{1}{p}-1)}$, with
$\lim_{n\rightarrow\infty} v_n = 1$.

For any $p$, $v_n -\lim_{n\rightarrow\infty} v_n \approx \frac{(1/p)-1}{n}$. This is a much slower rate of convergence. This greatly affects the box-counting dimension of the parametrized SNS's mixed-state presentation. In fact, $\text{dim}_0(Y)$ is nonzero (and roughly $1/2$) only when $p=q$; see Fig.~\ref{fig:SNS}(bottom). Additionally, for the parametrized SNS, causal states act as a counter of the number of $0$'s since last $1$. So, we can think of the predictive distortion at coarse-graining $\epsilon$ as $d(\Rep_{\epsilon}) \approx \EE-\EE(N_{\epsilon})$, which decreases exponentially quickly with $N_{\epsilon}$ rather than algebraically under weak conditions. (See, for example, Ref. \cite{travers2014exponential} and references therein.) Alternatively, from Eq.~(\ref{eq:d_L_ubound}), we note that $\sum_{i=0}^n \pi(i)$ decreases exponentially for the parametrized SNS; see Ref.  \cite{Marz14b} for details.
%Together, we find that the scaling of the number of features with desired
%distortion $d$ for countably infinite \eMs\ is very different from either
%finite or uncountably infinite \eMs.
From Fig.~\ref{fig:SNS}, we see that:
\begin{align*}
N_{\epsilon} \sim \begin{cases} \sqrt{1/\epsilon} & p=q \\ \left(1/\epsilon\right)^2 & p\neq q\end{cases}
  ~.
\end{align*}

